# A multi-technique approach to understanding delithiation damage in LiCoO$_2$ thin films

E. Salagre[1], S. Quílez[1], R. de Benito[1], M. Jaafar[1,2], H. P. van der Meulen[3,4], E. Vasco[5], R. Cid[5,6,7], E. J. Fuller,[8] A.A. Talin[8], P. Segovia[1,2,4], E.G. Michel[1,2,4*], and C. Polop[1,2,4*]

[1] Dpto. Física Materia Condensada, Univ. Autónoma de Madrid, Spain

[2] IFIMAC (Condensed Matter Physics Center), Univ. Autónoma de Madrid, Spain

[3] Dpto. Física de Materiales, Univ. Autónoma de Madrid, Spain

[4] Instituto Universitario de Ciencia de Materiales Nicolás Cabrera, Univ. Autónoma de Madrid, Spain

[5] Instituto de Ciencia de Materiales de Madrid, Consejo Superior de Investigaciones Científicas, Spain

[6] BM25-SpLine (Spanish CRG Beamline) at the European Synchrotron (E.S.R.F.), Grenoble, France

[7] Centre for Cooperative Research on Alternative Energies (CIC energiGUNE), Basque Research and Technology Alliance (BRTA), Vitoria-Gasteiz, Spain

[8] Sandia National Laboratories, Livermore (CA), USA

We report on the delithiation of LiCoO$_2$ thin films using oxalic acid (C$_2$H$_2$O$_4$) with the goal of understanding the structural degradation of an insertion oxide associated with Li chemical extraction. Using a multi-technique approach that includes synchrotron radiation x-ray diffraction, scanning electron microscopy, micro Raman spectroscopy, photoelectron spectroscopy and conductive atomic force microscopy we reveal the balance between selective Li extraction and structural damage. We identify three different delithiation regimes, related to surface processes, bulk delithiation and damage generation. We find that only a fraction of the grains is affected by the delithiation process, which may create local inhomogeneities. The chemical route to Li extraction provides additional opportunities to investigate delithiation while avoiding the complications associated with electrolyte breakdown and could simplify in situ measurements.

Lithium-ion batteries (LIBs) are widely used in current consumer electronics, and their demand in electric and hybrid vehicles as also for powering the internet of the things (IoT) is expected to grow rapidly in the near future [1, 2]. The most common cathode material in LIBs for consumer electronics is lithium cobalt oxide ($Li_xCoO_2$, LCO). LCO crystallizes in the α-$NaFeO_2$ structure, a layered rock-salt structure made up of alternating layers of $CoO_2$ and Li [3]. The layered structure allows a high mobility for Li atoms that enables LCO to reversibly intercalate Li. This feature is behind the excellent properties of LCO as a Li storage material in rechargeable LIBs, where the Li content can change by up to ~40% during the battery charging/discharging in a standard electrochemical process [4, 5]. Despite decades of research, questions remain regarding damage associated with Li insertion and extraction, reflecting the complex and interrelated processes involving electronic, chemical and structural aspects of the cathode materials [6]. Electrochemical Li extraction is frequently used to produce LCO films with different Li content, which can then be subjected to various types of ex situ analysis. While attractive in many respects, electrochemical Li extraction can also present numerous challenges including electrochemical side reactions related to electrolyte degradation and strong dependence of reaction rates on local electronic conductivity. Alternatively, chemical delithiation methods are significantly simpler experimentally, can be tuned to remove precise Li content using reactant concentration, and are independent of the local cathode electronic conductivity [7, 8]. A dilute acid solution in contact with LCO extracts Li, but it may also extract Co and induce an irreversible damage in the structure [9].

Previously, it was found that structural effects produced by both electrochemical and chemical delithiation methods are similar, i.e. the lattice constants and other structural parameters of chemically delithiated LCO are in good agreement with those of LCO delithiated by electrochemical methods for a significant delithiation range [5]. Nonetheless, other aspects such as the activity for the oxygen evolution reaction depend on the Li extraction method used [10]. Furthermore, chemical delithiation is in general studied for LCO particle-based electrodes, which can be challenging to interpret due to their heterogeneity and the presence of organic binders and electronic conductivity additives. Thin-films on the other hand, do not contain such additives and are much more amenable to investigation using traditional surface science and scanning probe methods. Among the different acids employed in chemical delithiation, oxalic acid ($C_2H_2O_4$) has been reported to be the most selective for removing Li ions from LCO cathodes with minimal extraction of Co [9]. Zheng et al. have proposed the following reaction to describe the process [11],

$$5H_2C_2O_{4(l)} + 2LiCoO_{2(s)} \rightarrow 2LiHC_2O_{4\ (l)} + 2CoC_2O_{4(s)} + 4H_2O_{(l)} + 2CO_{2(g)} \qquad (1)$$

This reaction produces LiHC$_2$O$_4$, which is moderately soluble in H$_2$O, and Co oxalate CoC$_2$O$_4$, which precipitates. However, there is no agreement on the effectiveness of this procedure and its consequences on the structural integrity of the LCO lattice [12].

In this work, we investigate the processes of chemical delithiation and associated structural defect formation in cathodes of LCO thin films with the goal of understanding the degradation processes involved in the delithiation and in order to correlate results obtained using experimental techniques sensitive to different properties of the LCO thin layer. In this approach we analyze strictly the role of oxalic acid and eliminate the side effects related to the presence of mixed electrodes (e.g. including PVDF, conducting carbon, etc). Thin films of LiCoO$_2$ grown by magnetron sputtering are exposed *ex-situ* to oxalic acid (C$_2$H$_2$O$_4$). The processes of structural changes and damage to the LCO films are investigated using synchrotron radiation powder XRD (x-ray diffraction), XPS (X-ray Photoelectron Spectroscopy), SEM (Scanning Electron Microscopy), micro Raman spectroscopy and C-AFM (Conductive Atomic Force Microscopy). We observe three different regimes related to surface processes, bulk delithiation and damage generation. A different balance between Li extraction and formation of structural damage characterizes each regime during the chemical delithiation of LCO.

**RESULTS AND DISCUSSION**

The reaction of oxalic acid with LCO films defines three delithiation regimes for the concentration and exposure times probed, as shown in Fig. 1. Low concentration (≤ 10$^{-3}$ M) and short exposure times (≤ 10 min) correspond to Regime I, defined by the presence of surface processes only. In Regime II (moderate concentrations of 10$^{-2}$ M and times below 70 min) bulk delithiation is observed. Larger exposure times (≥ 100 min) at concentrations of 10$^{-2}$ M and larger correspond to Regime III, where lattice damages are observed. These damages provide us with a measure of the loss of selectivity of the leaching reaction. The features of each regime are described in more detail in the following.

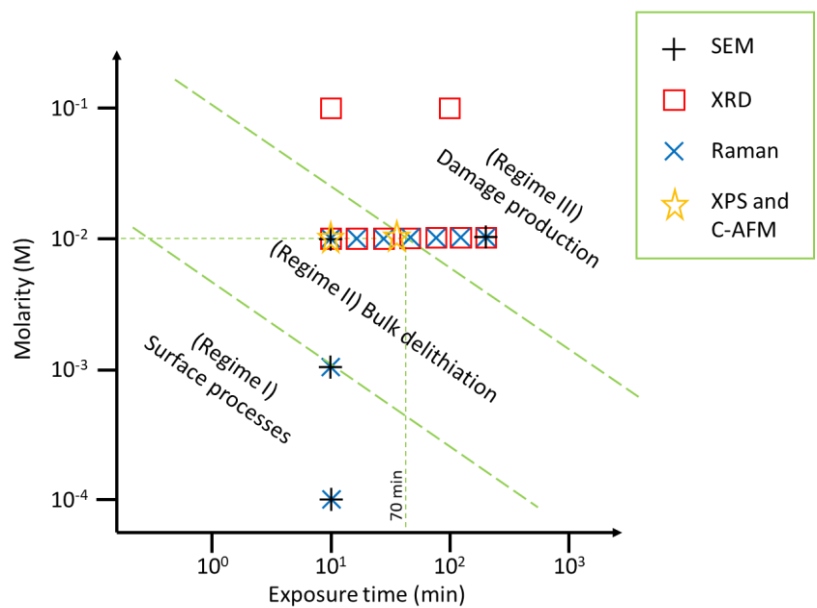

**Figure 1.** Overview of the different chemical delithiation regimes of LCO thin films with a diluted solution of oxalic acid. Symbols represent the parameters of acid concentration and exposure time that have been used in the experiments. Dashed green lines tentatively separate the regions of the parameter space of the different regimes.

SEM provides an overview of the LCO film morphology in different delithiation regimes as a function of the exposure conditions. The evolution of delithiation-induced features is summarized in Fig. 2. The as-grown film shows a typical polycrystalline morphology characterized by dispersed faceted grains with an average size of 60 nm (Fig. 2(a)). A mild exposure to oxalic acid with a molar concentration of $10^{-4}$ M induces changes in the surface of the grains, which coalesce into bundles (regions enclosed by a dashed green contour in Fig. 2(b) and (c)). These surface structures grow with the exposure time, reaching an average size of 90 nm for 10 min-exposure. The size increase is due mainly to the grouping of grains, rather than to their slight coarsening, as demonstrated by a statistical analysis using self-correlation (ACO) and power spectral density (PSD) functions (Fig. S2 in Suppl. Info.). The larger surface structures cause both the film roughness and porosity to decrease slightly, as shown in Fig. 2(f). These changes are likely due to the surface delithiation of the grains (Regime I in Fig. 1), as this process will affect first the surface morphology of the grains. The separation between surface roughness and bulk porosity coming to the surface is done by fitting the height distribution of the calibrated images using an outer Gaussian function centered on the average thickness (i.e., the mode of the height distribution) and an asymmetric background that corresponds to the tail of the distribution of emerging porosity (Fig. S3 in Suppl. Info.). We have ruled out that the surface phenomenon inducing an initial decrease in both porosity and roughness is an effect of distilled water used as a solvent. Exposure of LCO films to distilled water without acid produces a grain dispersion rather than a bundling, which results in a contrary behavior of roughness, i.e., roughness increases.

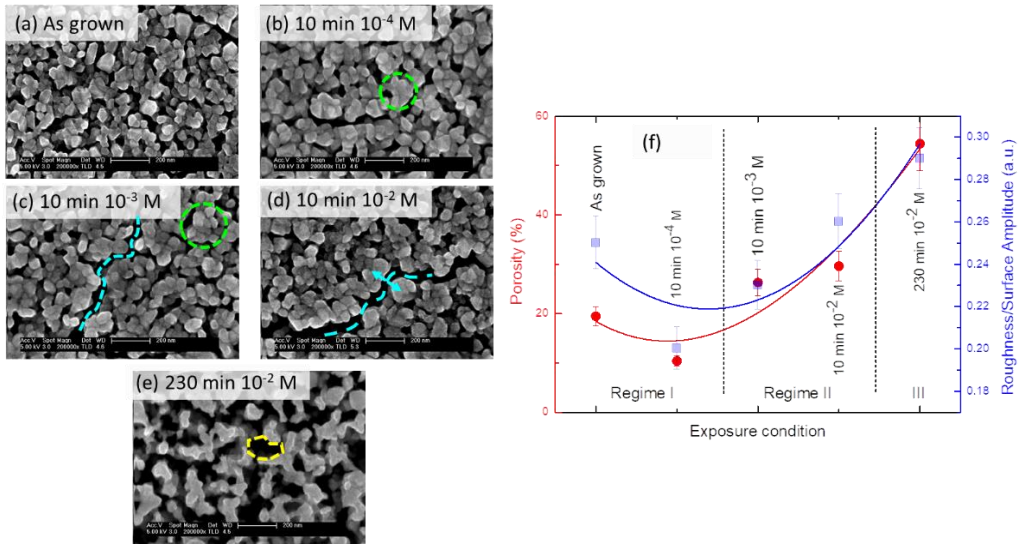

**Figure 2.** Evolution of the LCO thin film structure imaged by SEM for different exposure conditions to oxalic acid. Scale bar in (a-e) corresponds to 200 nm. Regions enclosed by a dashed green contour in (b) and (c) correspond to bundles of coalescing grains. Dashed light blue curves in (c) and (d) depict grooves. The blue arrow in (d) highlights the widening of the grooves. The dashed yellow section in (e) highlights the cross-section of a channel. (f) Roughness and film porosity for different exposure conditions. Curves serve as a visual guide.

As the acid concentration is increased up to $10^{-3}$ M, we begin to observe bulk delithiation. The coexistence of both surface and bulk processes produces bundles of grains separated by deep grooves (dashed light blue curve in Fig. 2(c)). This increases the roughness and porosity of the LCO films. Longer exposure times at lower concentrations and/or a further increase in the exposure concentration ($10^{-2}$ M) enhance the bulk delithiation (Regime II in Fig. 1), widening the grooves, as shown by the dashed light blue curve and blue arrow in Fig. 2(d).

Longer exposures and higher acid concentrations substantially increase the degree of delithiation and the observed damage (Regime III). The increased porosity gives rise to a network of channels (dashed yellow section in Fig. 2(e)), resulting in a high-specific surface area morphology. In addition to homogeneous damage throughout the film (regions A in Fig. 3), we also observe highly localized features such as expelled material and pits (regions B).

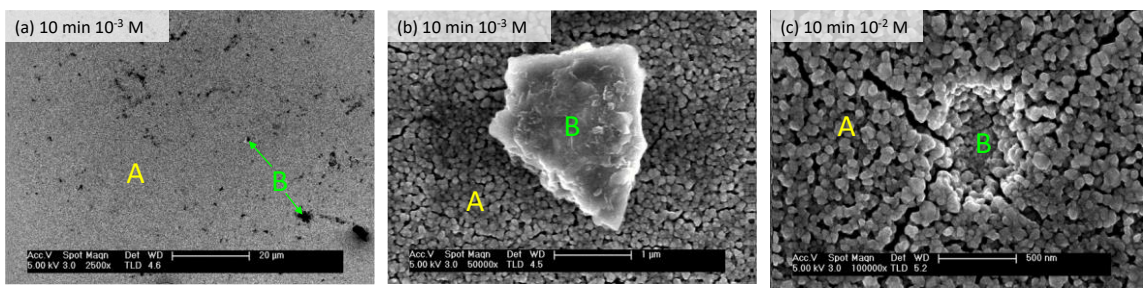

**Figure 3.** Homogeneous and heterogeneous damages (regions labelled A and B, respectively) in LCO thin films caused by exposures to oxalic acid. Scale bars correspond to: (a) 20 µm, (b) 1 µm and (c) 500 nm.

We have performed grazing incidence XRD in-plane measurements for as-grown samples and after successive oxalic acid exposures, in order to investigate the structural changes of LCO upon delithiation. We have monitored the main x-ray reflections for LCO, specifically (003), (104), (101), (10-5) and (107) during the process. The structure of as-grown LCO is rhombohedral (R-3m space group), but is typically described with a simpler hexagonal structure, usually labelled hex-I. Previous work has shown that delithiation causes structural phase transitions [4, 16, 17]. The hex-I phase of as-grown LCO is observed for Li concentrations in the range $0.95 \leq x \leq 1$. In the range $0.75 \leq x \leq 0.95$ there is phase coexistence between the hex-I phase and a second phase (called hex-II) with lower Li contents. The hex-II phase is similar to hex-I, but has a larger $c$ lattice constant. In the range $0.50 \leq x \leq 0.75$, only the hex-II phase is observed. Near x=0.5, a new phase of monoclinic symmetry appears. The hex-I phase corresponding to as-grown LCO is electrically insulating, while the hex-II phase is metallic, so that an insulator-metal phase transition is observed when the Li contents decreases [16, 18, 19]. Fig. 4 shows an XRD spectrum characteristic of LCO. The x-ray reflections are indexed according to a hexagonal lattice. Several residual reflections from the Pt layer underneath the LCO film and the Si substrate are also detected. XRD spectra after 10 min and after 150 min of exposure to $10^{-2}$ M oxalic acid are also shown, corresponding to Regime II (bulk delithiation) and Regime III (damage), respectively. As the film density decreases due to exposure to oxalic acid, more x-rays are able to penetrate to the substrate, and a significant increase of Pt and substrate peaks is observed for 150 min exposure time. The formation of pits in the LCO film also contribute to this effect. Note that no signal of crystalline Co oxalate is detected [11, 20, 21] (see Fig. S4 in Suppl. Info.).

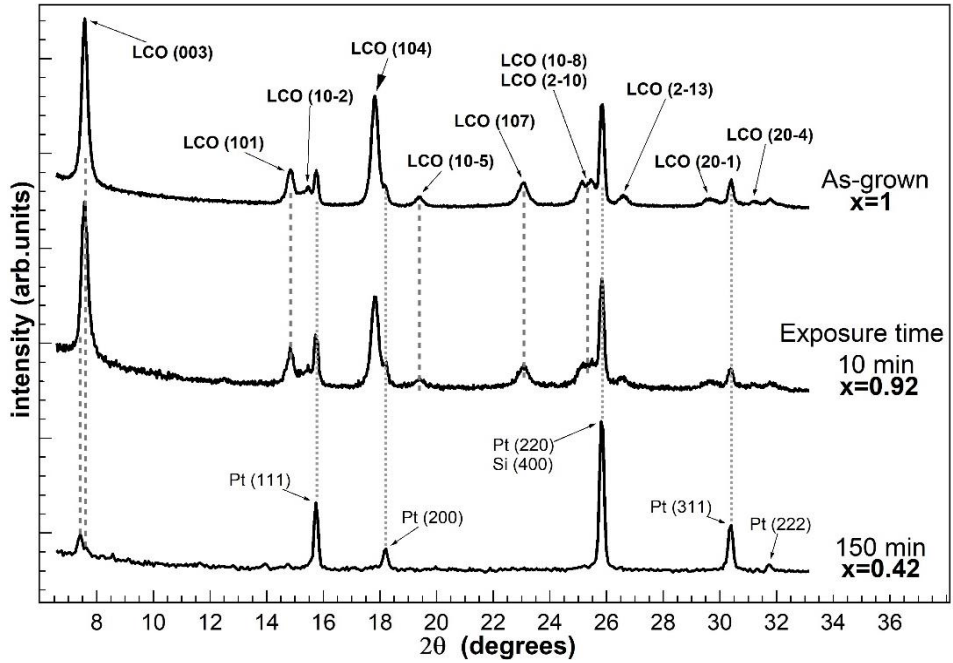

**Figure 4.** X-ray powder diffraction spectra corresponding to the as-grown LCO structure (top) and to samples delithiated 10 min (center) and 150 min (bottom) in $10^{-2}$ M oxalic acid. Reflections are indexed in the hexagonal structure. Residual reflections from the Pt layer underneath the LCO film and the Si substrate are detected.

Fig. 5(a) shows the evolution of the (003) reflection as a function of exposure time to oxalic acid ($10^{-2}$ M). The (003) reflection probes directly the $c$ lattice constant of the hexagonal structure. Li deintercalation produces a splitting of the (003) peak, due to the appearance of the hex-II structure with a larger $c$ lattice constant. The coexistence of the two structures hex-I and hex-II is in agreement with previous findings [23-22]. Fig. 5(b) shows the deconvolution of selected (003) reflections. Two dominant peaks are required to reproduce the line profile, one corresponding to the hex-I phase (blue) and one from the hex-II phase (orange). Two minor additional contributions have to be added, one corresponding to residual $Co_3O_4$ (green) and another one to account for strongly delithiated areas in the sample (magenta), due to local inhomogeneity [23]. The evolution of the dominant hex-I and hex-II components follows the expected behavior, from a dominant hex-I peak for the as-grown sample, to a dominant hex-II after strong delithiation. For each exposure time, the relative intensities of each component are represented as percentage of the total intensity of the (003) reflection for that particular time in Fig. 5(c). Note that the decrease of hex-I is related to the concomitant increase of hex-II and of the residual Co oxide signal. The data corresponding to exposure times below 70 min (x ≈ 0.6) correspond to regime II (bulk delithiation), while data above 70 min correspond to regime III (damage), which is characterized in XRD data by the increase of signal coming from Co oxide and substrate peaks (Pt, Si), due to the formation of significant pits and channels in the LCO film as imaged by SEM.

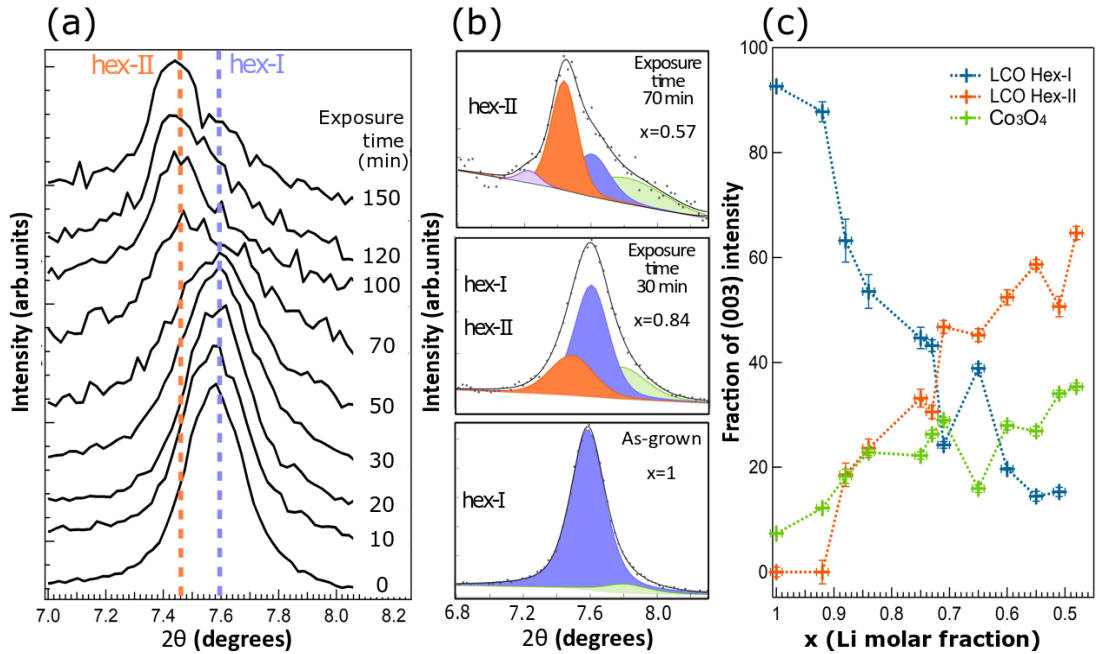

**Figure 5.** (a) Evolution of the (003) reflection in grazing incidence powder diffraction for increasing exposure times at $10^{-2}$ M concentration. Note the splitting of the peak due to the coexistence of the hex-I and hex-II structures with different c lattice parameter and the shift of the maximum to lower angles as the exposure time increases. (b) Deconvolution of selected peaks corresponding to the (003) reflection including hex-I (blue), hex-II (orange), $Co_3O_4$ (green), and residual strongly delithiated areas (magenta). (c) Intensity of the different components (hex-I, hex-II and $Co_3O_4$) in the (003) reflection as a function of Li molar fraction. For each exposure time, the relative intensities of components are normalized with respect to the total intensity of the (003) reflection for that particular exposure time.

Fig. 6 (a) displays the values of the *c* lattice parameter as a function of delithiation time and Li molar fraction obtained from the data in Fig. 5. The results are compared to measurements of the *c* lattice parameter for single-crystalline samples of $Li_xCoO_2$ of well-defined composition [24]. There is an overall good agreement between both sets of data, showing that the variation in lattice parameter *c* is consistent with the coexistence of hex-I and hex-II phases region for $0.75 < x < 0.95$. It is important to note that the chemical delithiation results in a heterogeneous phase transformation. This is deduced from Fig. 5, as the hex-I component persists for x values lower than reported (down to 0.5 instead of 0.75). Another clue leading to this conclusion is that the line profile of the (003) peak corresponding to a large delithiation (x=0.57, Fig. 5(b), bottom peak), shows a shoulder on its low angle side, indicating that some sample areas have lower lithium concentration than average. Overall, the values for the parameter *c* of the hex-II phase are slightly higher than reported, probably due to this heterogeneous Li deintercalation.

Fig. 6 (b) shows the intensity of different x-ray reflections as a function of exposure time and Li molar fraction, as determined from the (003) peak position. The intensities are normalized to the incident radiation intensity. Fig. 6(b) highlights the evolution from bulk delithiation towards damage. An almost linear decrease of the intensities is observed as a function of the exposure time in the range for 0.5 < x < 1. In this range, the bulk delithiation takes place, as deduced from the analysis of the *c* lattice constant. Interestingly, the evolution of the lattice parameters of both the phase hex-I and hex-II follows closely the behavior found for electrochemical delithiation or for single crystals [4, 24]. Damage is gradually introduced, as evidenced by the drop in intensity for all reflections (Fig. 6 (b)), which become residual with respect to the substrate contribution above 60-70 min (Regime III). The damage is also evidenced from the growth of the relative importance of Co oxide peaks (Fig. 5). These results agree with the morphological description obtained by SEM.

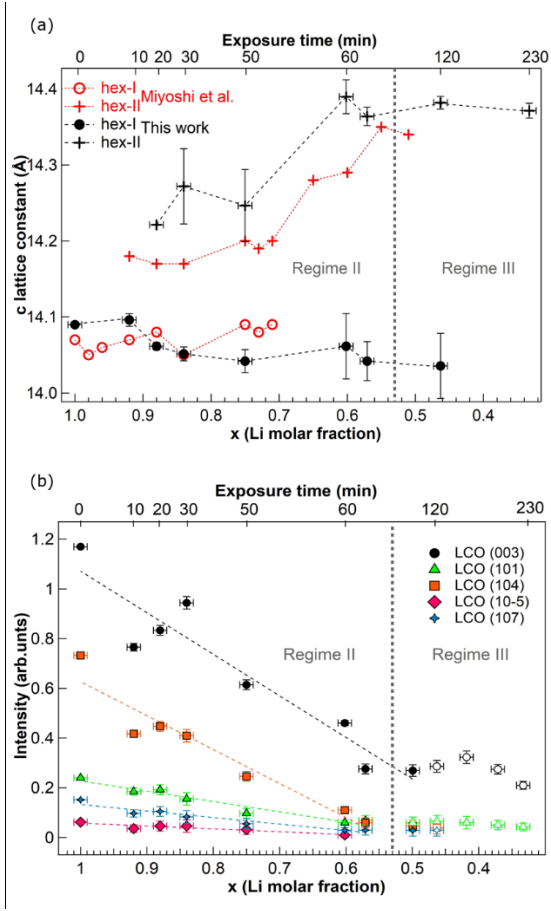

**Figure 6.** Dependences of (a) lattice parameter c for $Li_xCoO_2$ and (b) the intensities of x-ray reflections on the exposure time at $10^{-2}$ M concentration (top axis) and Li molar fraction (bottom axis). The appearance of two coexisting c values due to phases hex I and hex II are compared to literature values reported in Ref. [24].

Fig. 7 shows a series of Raman spectra obtained for increasing exposure times at $10^{-2}$ M oxalic acid solution. A typical Si substrate coated only by a Pt/Ti buffer layer was measured as well. No signal was observed in the analyzed range, which demonstrates that Pt/Ti completely shields the 520 cm$^{-1}$ Si signal. The two most prominent peaks in Fig. 7 (a) correspond to the A1g and Eg modes of LCO. No changes in the Raman spectra are visible in Regime I, corresponding to lower acid concentrations and shorter exposure times (not shown). In Regime II (up to 70 minutes of exposure time at $10^{-2}$ M), the overall intensity of the spectra slowly increases and a broadening of both the Eg (485 cm$^{-1}$) and the A1g (593 cm$^{-1}$) peaks is observed at their low-energy side, as illustrated in Fig. 7(b) for the A1g peak. This is ascribed to the gradual transformation of LCO from the hex-I to the hex-II phase, the latter having the Raman peaks at slightly lower energies [25, 26]. A minor contribution from residual $Co_3O_4$ is also detected in Regime II (exposure times below 70 min). For exposure times beyond 70 minutes, corresponding to Regime III, peaks of $Co_3O_4$ increase as an indication that damage is introduced in the sample. The damage is also reflected in the higher contribution of the Raman signal in-between the peaks, which is a consequence of the appearance of structural disorder in the material. No Raman signal of Co oxalate has been detected, expected at a value of 783 cm$^{-1}$ [27].

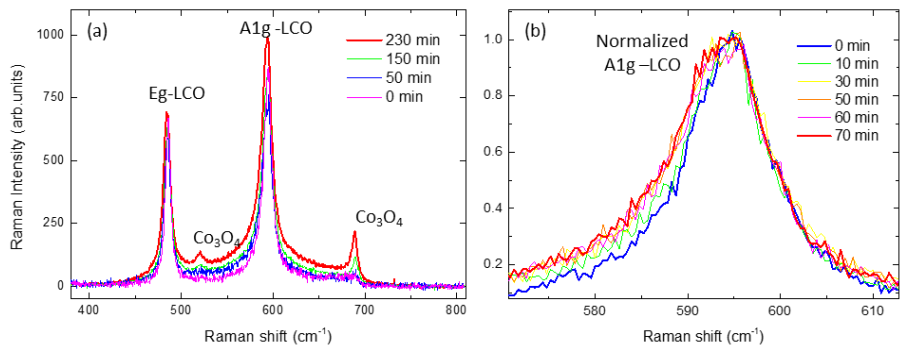

**Figure 7.** (a) Evolution of the Raman spectra for various exposure times at a $10^{-2}$ M solution of oxalic acid. (b) Evolution of the A1g-LCO peak (normalized to the height) for exposure times up to 70 minutes (corresponding to Regime II). The broadening of the left side is due to the appearance of the hex-II phase of LCO.

Fig. 8 represents the Raman intensity vs. exposure time. The insert in Fig. 8(a) shows that the total integrated intensity stays constant in Regime I (exposure times of 10 min at $10^{-4}$ and $10^{-3}$ M acid concentration). In turn, the total integrated intensity grows continuously during Regime II and III (Fig. 8(a)). This enhancement in Raman efficiency can be explained both by an increase in the surface to volume ratio due to the porosity evolution as observed by SEM and by the change in the bandgap of the material as the Li-concentration diminishes [28]. The Regimes II and III are recognized in Fig. 8 (b) and (c), where the relative integrated contribution of the Eg

Raman peak (summing up the two phases) is constant up to 70 minutes and decreases afterwards. The relative integrated contribution of the $Co_3O_4$ peak at 689 cm$^{-1}$ shows the opposite behavior, i.e. an increase in Regime III when the damage sets in, in agreement with XRD results.

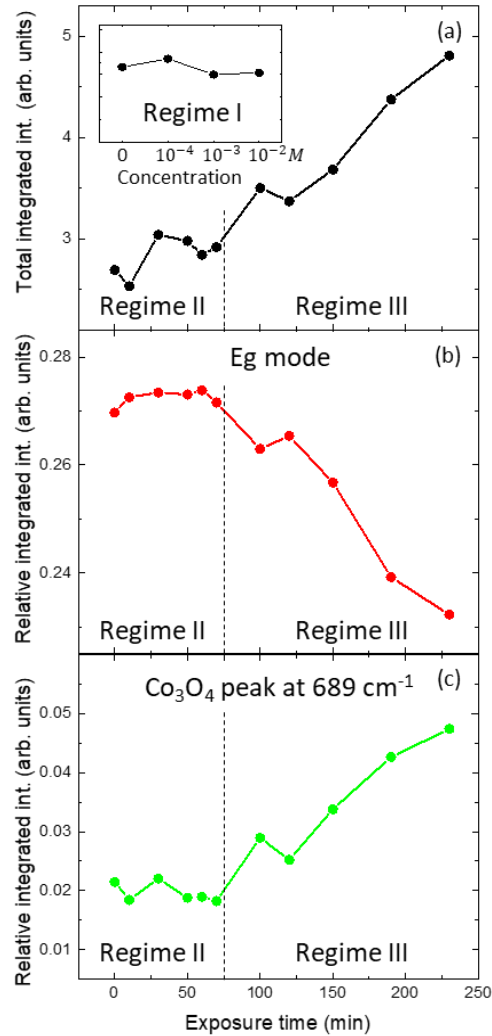

**Figure 8.** Evolution with exposure time at a 10$^{-2}$ M acid concentration of: (a) the total integrated Raman spectrum, (b) the relative integrated contribution of the Eg Raman peak (summing up the 2 phases), and (c) the relative integrated contribution of the $Co_3O_4$ peak at 689 cm$^{-1}$. Regimes II (below 70 min.) and III (above 70 min.) are clearly distinguishable in (b) and (c). The insert in (a) represents the total integrated Raman spectrum in Regime I, acid exposure times of 10 minutes at the lower concentrations.

An interesting feature in the Raman spectra is the remarkable persistence of the hex-I phase until the maximum exposure time. In the XRD data both phases are observed until a Li molar fraction of at least 0.45 (Fig. 6), corresponding to a delithiation time of 120 min. The relative evolution of the intensities of the two phases is also different in both techniques. In the Raman spectra the phase II signal is always lower than that of phase I, while in XRD experiments the signal of phase II replaces the signal of phase I. On one hand, the persistence of the hex-I phase beyond 0.75 molar fraction indicates that the sample is not homogeneous after the chemical delithiation, in agreement with the observation by XRD of regions further impoverished in Li (Fig. 5b, top). On the other hand, we may expect that the Raman measurement is more sensitive to the surface layer. This is because the optical penetration depth is reduced by the presence of the metallic hex-II phase as well as by the increase in absorption because of an approaching resonance when the Li-concentration diminishes (the latter also can be responsible for the net increase in the Raman signal) [28]. The persistence of the hex-I phase is in contrast with results of Raman spectra taken in an electrochemical delithiation process, where the Raman signal of the hex-I phase disappears as the Li-content falls below 0.8 [26]. The reason for the different behavior is presently unknown but may be due to how the LCO material is prepared and the uniformity of delithiation process.

In order to corroborate some of the results described above, complementary measurements by XPS and C-AFM were made in the as-grown sample and compared to results obtained in a delithiated sample in Regime II.

Full range XPS spectra are measured to check the sample composition and cleanliness. Fig. 9 shows XPS spectra corresponding to Co 3p and Li 1s, which are used to determine the Li contents for the as-grown sample and for a chemically delithiated sample (65 min in $10^{-2}$ M oxalic acid corresponding to Regime II). The core levels are normalized to the maximum of Co 3p to highlight the changes. Co 3p is a doublet, but the splitting of the components is small and is not observed with our experimental resolution [29]. The peak presents an asymmetric shape as described in Ref. [30]. The analysis of the Li 1s core level is hindered by its low cross section for the photon energy used. The peak is difficult to observe, especially in delithiated samples. Nonetheless, its change upon delithiation is clear. Comparing with published results [23], the observed Li 1s and Co 3p peak ratios together with the decrease of Li 1s height correspond to a value of x=0.5 for the delithiated sample, while for the as-grown sample the stoichiometry is close to x=1, as expected. The result demonstrates that oxalic acid at low concentrations is able to partially remove Li from LCO without imparting structural damage. Therefore, the chemical reaction for Li extraction in Regime I can be written simply as

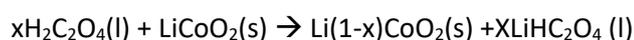

$xH_2C_2O_4(l) + LiCoO_2(s) \rightarrow Li_{(1-x)}CoO_2(s) + XLiHC_2O_4(l)$

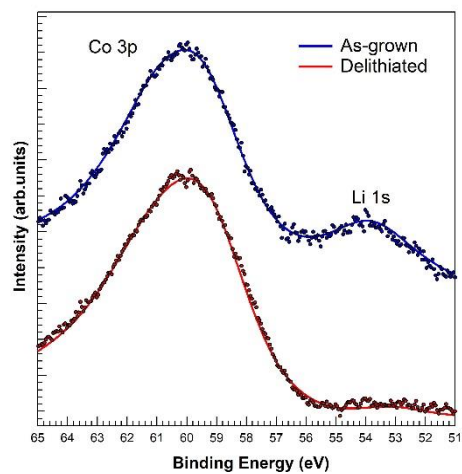

**Figure 9.** XPS Co 3p and Li 1s regions for the as-grown (top) and chemically delithiated (65 min at 10$^{-2}$ M oxalic acid, bottom) samples. Dots are experimental data and the line serves as a guide to the view.

Previous studies have shown the influence of the delithiation process in the electronic conductivity of LCO. It is well known that Li$_x$CoO$_2$ is insulating for $x \geq 0.95$, whereas a metallic conductivity appears for $0.35 \leq x \leq 0.95$ [31]. Milewska and coworkers report an enhancement of the electrical conductivity of 6 orders of magnitude upon electrochemical delithiation [18]. By comparing their results with other reports in the literature, they conclude that the electrical properties of thin LCO films differ from data obtained for powder LCO in terms of electrical conductivity. Moreover, a comparison between works using different preparation methods of LCO, which implies dissimilar structures of the material, may yield different behaviors [18].

Fig. 10 shows a qualitative comparison of the electrical conductivity (by C-AFM) between the same samples analyzed previously by XPS. Both samples present the typical grain morphology (Fig. 10(a, b) and Fig. S5 in Suppl. Info.) consistent with SEM. The electrical current map measured at a bias of -1.5 V in the as-grown sample, Fig. 10(c), shows a moderately heterogeneous spatial distribution of low conductivity, likely due to the morphology of the polycrystalline films. In turn, the current map in the delithiated sample, Fig. 10(d), shows a highly heterogeneous conductivity with enhanced regions, probably where delithiation is deeper. The increase in the heterogeneity, as revealed by the comparison in Fig 10(e) of the widths of the corresponding histograms, is due to two effects: (i) the heterogeneous structure inherent to the polycrystalline film, and (ii) the heterogeneity in the delithiation front, which are closely related to each other as suggested by the SEM measurements in Fig. 2. We have estimated typical conductivity values of the order of 10$^{-2}$ S cm$^{-1}$, from current maps and taking into account tip-sample contact areas and sample thickness. The conductivity values are in the range of those previously reported for electrochemically delithiated LCO [18]. However, we note that we are using two-point method measurements, so that series resistance could affect them.

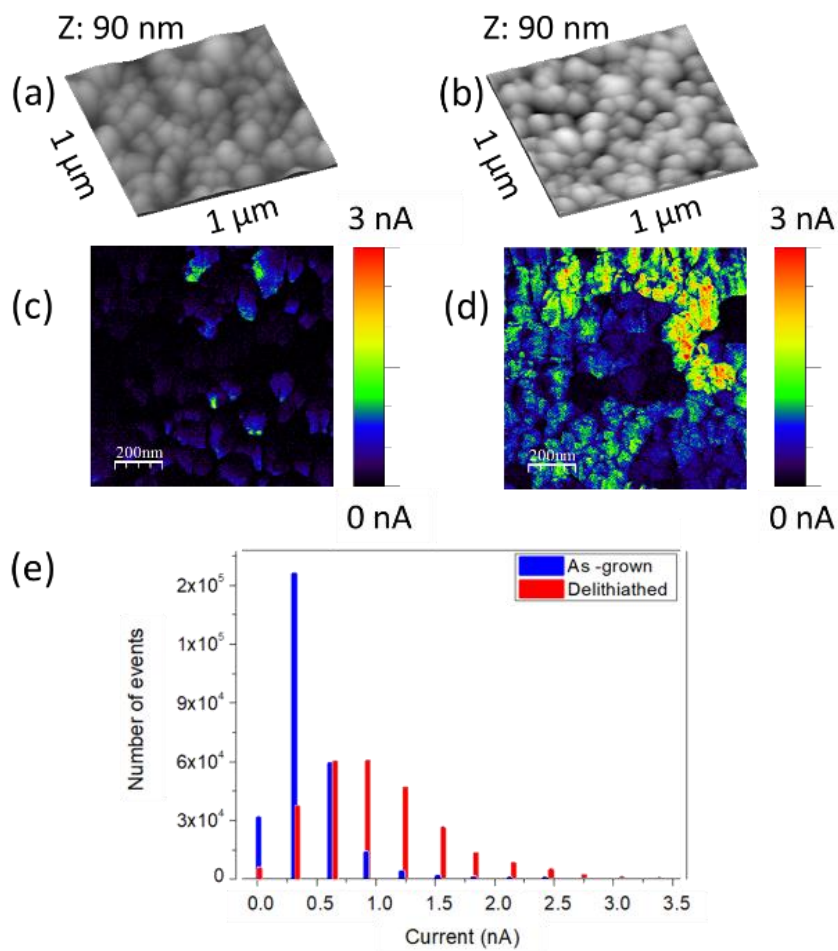

**Figure 10.** Topography images (a and b) and current maps (c and d) of the as-grown and delithiated samples (65 min at $10^{-2}$ M oxalic acid), respectively. Images size 1 μm x 1 μm. (e) Histogram of the current maps. Both experiments are performed with the same tip and parameters: force=150 nN and bias voltage=-1.5 V.

The multi-technique approach by using SEM, XRD, Raman, XPS and C-AFM reveals the existence of three regimes of chemical delithiation of the LCO film by exposure to oxalic acid, depending on the acid molar concentration and exposure time conditions, as shown in Fig. 1. The regimes differ in the morphology, structures, and chemical composition of the LCO film that result from the leaching reaction. Regime I corresponds to a mild surface delithiation causing a bundling of grains. Regime II is effective to delithiate the bulk of the LCO film, as evidenced by the changes in Li concentration, the appearance of phase hex-II and the evolution of the lattice constants of hex-I and hex-II phases. All these features mimic the behavior of LCO upon electrochemical delithiation [4, 16, 17, 18], and that detected in single crystalline samples [24]. Additionally, in regime II, groove formation starts to occur. Further acid exposure (Regime III) corresponds to increased structural degradation and amorphization of the LCO film, which produces high-specific surface area structures. However, rather than abrupt crossover between regimes, ranges of coexistence are observed, illustrating the spatial heterogeneity of the reaction. These findings support that the chemical delithiation reaction takes place predominantly at the surface of the LCO grains and is limited by the reduced mobility of Li in the granular volume of the film. As a result, regions with different Li stoichiometries can coexist. Probably, $Co_3O_4$ segregated to the inter-grain regions acts as a barrier to the diffusion of Li.

The suspicion that precipitated Co oxalate could contribute to this barrier is ruled out, since no distinctive signs of this compound were observed by XRD and Raman. Thus, our results do not support that the reaction of equation (1) describes the process. Instead, our results demonstrate that oxalic acid used in low molar concentrations ($10^{-4}$ ... $10^{-2}$ M) is a highly selective leaching agent for delithiation preserving the LCO structure.

**CONCLUSIONS**

Mild chemical delithiation of LCO thin films with oxalic acid is observed for a range of moderate acid concentrations and exposure times (Regimens I and II). Under these conditions, Li is selectively extracted after exposing LCO films to oxalic acid and the resulting $Li_xCoO_2$ has structural lattice constants and electrical properties similar to electrochemically delithiated LCO, albeit in a more heterogeneous manner. Deeper delithiation corresponding to increased acid concentrations and longer exposure times (Regimen III), results in a less selective Li extraction and irreversible structural damage of the LCO films. The chemical route to Li extraction provides additional opportunities to investigate delithiation while avoiding the complications associated with electrolyte breakdown and could simplify in situ measurements.

**EXPERIMENTAL METHODS**

**LCO film growth**

The samples are prepared as follows. A 10 nm Pt/Ti buffer layer is first deposited onto a Si(001) substrate capped with a 100 nm thick thermal $SiO_2$, using electron beam evaporation. LCO (nominal thickness ≈500 nm) is then deposited by magnetron sputtering in the same system, keeping the substrate at room temperature. The LCO target (Kurt J. Lesker) is sputtered with a 230 Watt plasma from a gas mixture of 3:1 Ar:$O_2$ at an static pressure of 16 mTorr. Following deposition, the LCO film is annealed in oxygen at 600 °C for 2 h in order to form the high-temperature (HT) LCO phase [13].

**Chemical delithiation**

Chemical extraction of Li from the polycrystalline LCO thin film is performed using a solution of oxalic acid ($C_2H_2O_4$) in distilled water and variable exposure times. Nominally, a $10^{-2}$ M concentration is used, but lower concentrations are also employed. As the samples are compact polycrystalline thin films of LCO, we used the drop method for the delithiation process. A drop of the solution is deposited at room conditions on top of the sample, which is maintained by surface tension of the liquid covering the entire sample surface. After exposure, the sample surface is rinsed with distilled water and analyzed immediately after.

**Characterization**

SEM is performed using a field-emission Philips XL-30 microscope equipped with a through lens detector (TLD). The measurements are carried out in a vacuum of $10^{-6}$ mbar, at an acceleration voltage of 5 kV, with magnifications ranged between 25 k and 200 k.

XRD measurements are done in BM25 beamline of the ESRF using a six-circle diffractometer in vertical geometry [14] equipped with a 2x2 Maxipix detector on the diffractometer arm (Fig. S1 in Supplementary Information). Three circles are dedicated to the sample motion ($\chi$, $\phi$ and $\theta$), two circles are dedicated to the detector motion ($\Gamma$, $\delta$), and a rotation ($\mu$) coupled to the sample and detector motion is present. The $\theta$-circle rotates the sample around the direction normal to the sample surface and the ($\chi$, $\phi$)-circles are used only for the sample alignment. The $\mu$-circle sets the beam incidence angle, which can be varied between 0.0 and 5.0 degrees. The $\Gamma$-detector circle performs out-of-plane measurements while the $\delta$-circle, when $\Gamma=0$, performs in-plane measurements. The diffractometer sphere of confusion for all axis circles is 50 µm. A photon energy of 20 KeV is used for all measurements.

Micro Raman spectra are recorded in a backscattering geometry. A 532 nm excitation laser beam with power of 2 mW is focused to a 1.5 µm-diameter spot by a ×50 objective lens (NA = 0.73). The scattered light is collected by the same objective lens, passes through a 532-nm long-pass edge filter, is dispersed by a single-grating monochromator and is detected with a liquid nitrogen-cooled CCD camera.

XPS experiments are performed in an ultrahigh vacuum (UHV) chamber. Mg K$_\alpha$ radiation (non-monochromatic) is used to excite core level photoelectrons, which are detected with a Phoibos 150 hemispherical analyzer. Before performing the XPS experiments, the LCO sample is cleaned by annealing at 500 °C in oxygen atmosphere ($10^{-6}$ mbar). This procedure is required to remove a thin layer of surface contamination containing adventitious carbon and due to exposure to the atmosphere. The cleaning procedure is found to maintain the overall stoichiometry of the LCO sample. The Li content is measured by comparing the intensities of Li 1s and Co 3p core level peaks, which are very close to each other.

C-AFM is carried out using a Nanotec Electronica S.L. AFM tool and WSxM software [15]. We use cantilevers from Nanosensors (PtSi-NCH, PtSi coating, nominal resonance frequency of 330 kHz and spring constant of 42 Nm$^{-1}$) and from Budget Sensors (ElectriMulti75-G, CrPt coating, nominal resonance frequency of 75 kHz and spring constant of 2.8 Nm$^{-1}$). Topography and current images are taken simultaneously in contact mode at room conditions. The tip is used as a moving electrode with constant voltage applied to it while the specimen contacted by the bottom Pt/Ti electrode is grounded.

**REFERENCES**

[1]  N. Nitta, F. Wu, J. T. Lee, and G. Yushin, *Mater. Today* **18**, 252 (2015).

[2]  J. B. Goodenough and Y. Kim, *Chem. Mater.* **22**, 587 (2010).

[3]  K. Mizushima, P. C. Jones, P. J. Wiseman, and J. B. Goodenough, *Mater. Res. Bull.* **15**, 783 (1980).

[4]  J. N. Reimers and J. R. Dahn, *J. Electrochem. Soc.* **139**, 2091 (1992).

[5]  Y. Takahashi, N. Kijima, K. Dokko, M. Nishizawa, I. Uchida, and J. Akimoto, *J. Solid State Chem.* **180**, 313 (2007).

[6]  J. P. Pender *et al.*, Electrode Degradation in Lithium-Ion Batteries, *ACS Nano* **14**, 1243-1295 (2020).

[7]  J. Graetz, A. Hightower, C. C. Ahn, R. Yazami, P. Rez, and B. Fultz, *J. Phys. Chem.* B **106**, 1286 (2002).

[8]  R. Gupta and A. Manthiram, *J. Solid State Chem.* **121**, 483 (1996).

[9]  M. Aaltonen, C. Peng, B. P. Wilson, and M. Lundström, *Recycling* **2**, (2017).

[10] V. Augustyn and A. Manthiram, *J. Phys. Chem. Lett.* **6**, 3787 (2015).

[11] X. Zeng, J. Li, B. Shen, *Journal of Hazardous Materials* **295**, 112 (2015).

[12] M. H. Rodriguez, D. S. Suarez, E. G. Pinna, C. Zeballos, Patent, International Application No. PCT/IB2016/056189, Publication number WO/2017/064677, Method for the acid dissolution of LiCoO$_2$ contained in spent lithium-ion batteries (2017).

[13] S. Tintignac, R. Baddour-Hadjeana, J.P. Pereira-Ramosa, R. Salotb, *Electrochimica Acta* **60**, 121 (2012).

[14] J. Rubio-Zuazo, P. Ferrer, A. López, A. Gutiérrez-León, I. da Silva, and G.R. Castro, *Nucl. Instrum. Methods Phys.* **A716**, 23 (2013).

[15] I. Horcas, R. Fernández, J. M. Gómez-Rodríguez, J. Colchero, J. Gómez-Herrero, and A. M. Baro, *Rev. Sci. Instrum.* **78**, 013705 (2007).

[16] M. Menetrier, I. Saadoune, S. Levasseur, C. Delmas, *J. Mater. Chem.* **9**, 1135 (1999).

[17] G. G. Amatucci, J.M. Tarascon, L. Klein, *J. Electrochem. Soc.* **143**, 1114 (1996).

[18] A. Milewska, K. Świerczek, J. Tobola, F. Boudoire, Y. Hu, D. K. Bora, B. S. Mun, A. Braun, and J. Molenda, *Solid State Ionics* **263**, 110–118 (2014).

[19] C. A. Marianetti, G. Kotliar, G. Ceder, *Nat. Mater.* **3**, 627 (2004).

[20] R. Deyrieux, C. Berro, A. Peneloux, *Bulletin de la Societe Chimique de France* **1**, p25-34 (1973).

[21] L. Ren, P. Wang, Y. Han, C. Hu. B. Wei, *Chem. Phys. Lett.* **476**, 78-83 (2009).

[22] L. Dahéron, R. Dedryvère, H. Martinez, M. Ménétrier, C. Denage, C. Delmas, and D. Gonbeau, *Chem. Mater.* **20**, 583 (2008).

[23] F. Lin, I. M. Markus, D. Nordlund, T-C. Weng, M. D. Asta, H. L. Xin, and M. M. Doeff, *Nat. Comm.* **5**, 3529 (2014).

[24] K. Miyoshi, K. Manami, R. Sasai, S. Nishigori, and J. Takeuchi, *Phys. Rev. B* **98**, 195106 (2018).

[25] M. Inaba, Y. Iriyama, Z. Ogumi, Y. Todzuka, and A. Tasaka, *J. Raman Spectrosc.* **28**, 613 (1997).

[26] C. M. Julien and A. Mauger, *AIMS Mater. Sci.* **5**, 650 (2018).

[27] B. K. Pandey, A. Sukla, A. K. Sinha, and R. Gopal, *Materials Focus* **4**, 333 (2015).

[28] H. L. Liu, T. Y. Ou-Yang, H. H. Tsai, P. A. Lin, H. T. Jeng, G. J. Shu, and F. C. Chou, *New J. Phys.* **17**, 103004 (2015).

[29] A. Lebugle, U. Axelsson, R. Nyholm, and N. Mårtensson, *Phys. Scr.* **23**, 825 (1981).

[30] D. Gonbeau et al, *Chem. Mater.* **20**, 583–590 (2008).

[31] T. Motohashi, Y. Sugimoto, Y. Masubuchi, T. Sasagawa, W. Koshibae, T. Tohyama, H. Yamauchi, and S. Kikkawa, *Phys. Rev. B* **83**, 195128 (2011).

**ACKNOWLEDGMENTS**

This work has been supported by the Spanish MICINN (grant nr. FIS2017-82415-R, grant nr. MAT2017-83722-R, "María de Maeztu" Programme for Units of Excellence in R&D (CEX2018-000805-M)), within the framework of UE M-ERA.NET 2018 program under StressLIC Project (grant nr. PCI2019-103604 and PCI2019-103594) and by the Comunidad Autónoma de Madrid (grant nr. SI1/PJI/2019-00055, contract nr. PEJD-2019-PRE/IND-15769 and PEJ-2018-AI/IND-10072). The work at Sandia National Laboratories was supported by the Laboratory-Directed

**CONFLICTS OF INTEREST**

There are no conflicts of interest to declare.

**AUTHOR CONTRIBUTIONS**

E.S., R.B., R.C., P.S. and E.G.M. carried out and analyzed the XRD and XPS experiments. S. Q., M.J. and C.P. carried out and analyzed the C-AFM experiments. S.Q., R.B., H.P.M. and C.P. carried out and analyzed the Raman experiments. E. V. analyzed the SEM experiments. E.J.F. and A.A.T. prepared the samples. E.G.M. and C.P. devised and supervised the project. All authors discussed the results and contributed to the final manuscript.

# A multi-technique approach to understanding delithiation damage in LiCoO$_2$ thin films

**Supplementary Information**

E. Salagre[1], S. Quílez[1], R. de Benito[1], M. Jaafar[1,2], H. P. van der Meulen[3,4], E. Vasco[5], R. Cid[5,6,7], E. J. Fuller,[8] A.A. Talin[8], P. Segovia[1,2,4], E.G. Michel[1,2,4] *, and C. Polop[1,2,4] *

[1] Dpto. Física Materia Condensada, Univ. Autónoma de Madrid, Spain

[2] IFIMAC (Condensed Matter Physics Center), Univ. Autónoma de Madrid, Spain

[3] Dpto. Física de Materiales, Univ. Autónoma de Madrid, Spain

[4] Instituto Universitario de Ciencia de Materiales Nicolás Cabrera, Univ. Autónoma de Madrid, Spain

[5] Instituto de Ciencia de Materiales de Madrid, Consejo Superior de Investigaciones Científicas, Spain

[6] BM25-SpLine (Spanish CRG Beamline) at the European Synchrotron (ESRF), Grenoble, France

[7] Centre for Cooperative Research on Alternative Energies (CIC energiGUNE), Basque Research and Technology Alliance (BRTA), Vitoria-Gasteiz, Spain

[8] Sandia National Laboratories, Livermore (CA), USA

INDEX:

## 1) XRD experimental set-up

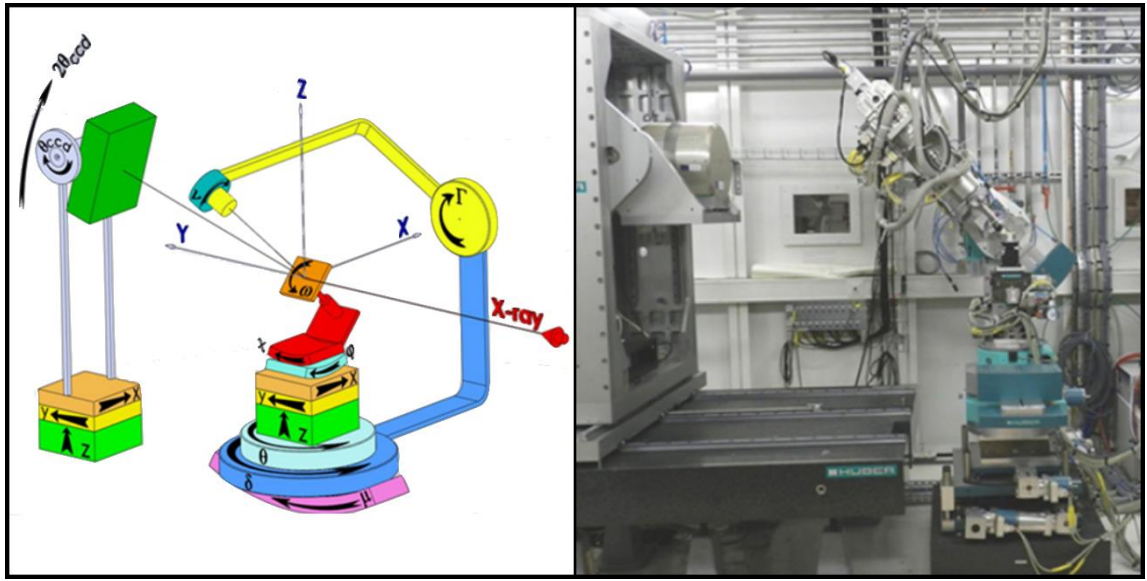

**Figure S1.** Left: x-ray diffraction set-up scheme. Right: experimental apparatus.

## 2) Statistics of structural properties by SEM

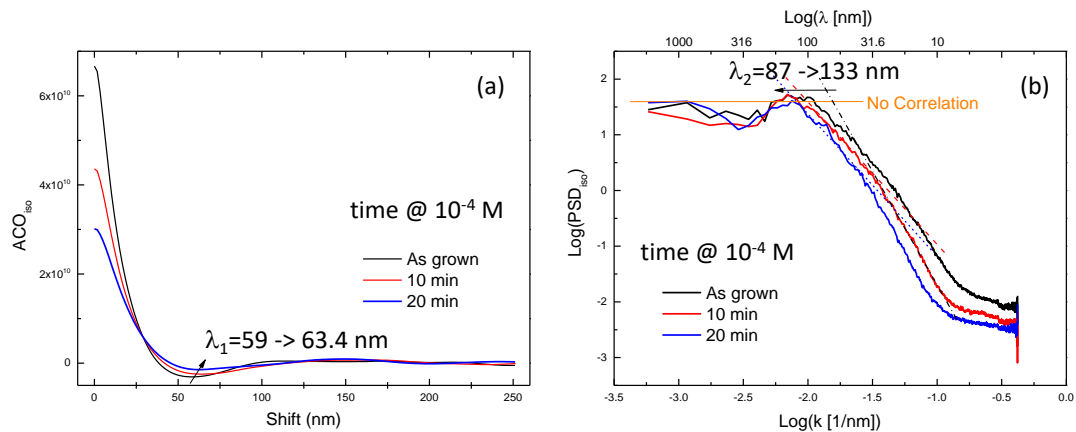

**Figure S2.** Statistical analysis of the increase in the size of the surface structures during Regime I ($10^{-4}$ M). Two lateral correlation lengths are identified and correlated to $\lambda_1$-grain size and $\lambda_2$-size of grain bundle. (a) Radial-averaged self-correlation functions (ACO$_{iso}$) for different exposure times reveal that $\lambda_1$, which is computed from their minima, increases slightly. On the other hand, (b) radial-averaged power spectra densities (PSD$_{iso}$) demonstrate that increase in $\lambda_2$ is higher. The increase is computed from the interceptions of their slopes with the "No correlation"-labelled line, which defines the length scale where any correlation is extinguished. This suggests that the increase in the structure size is due mainly to the grain coalescence into bundles, rather than grain growth.

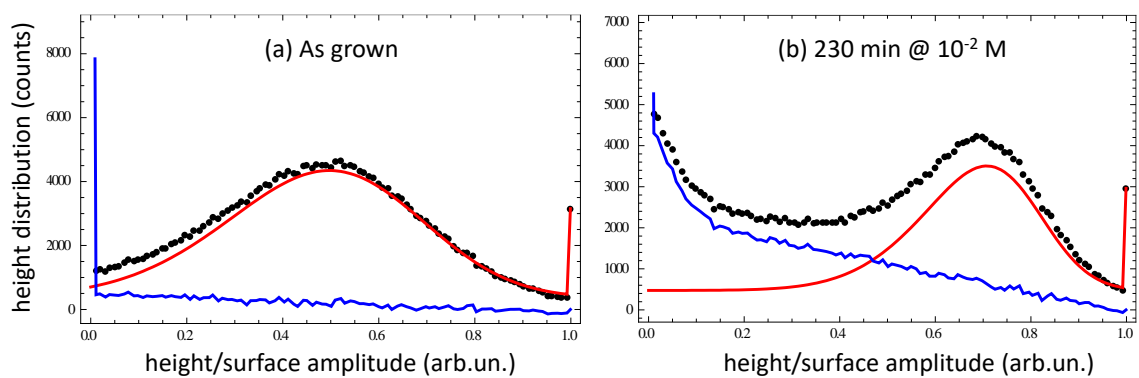

**Figure S3.** Evolution of the height distribution of the LCO thin film with the exposure conditions. Such distribution is calculated from calibrated SEM images. Symbols plot the experimental data and the curves denote their fit using a Gaussian function (red curve) to estimate the surface contribution and a background (blue curve) that corresponds to the tail coming to the surface of the bulk distribution of the porosity. The data on the ends accumulate the heights outside of the range of the measurements.

### 3) Full spectrum XRD including Co oxalate information

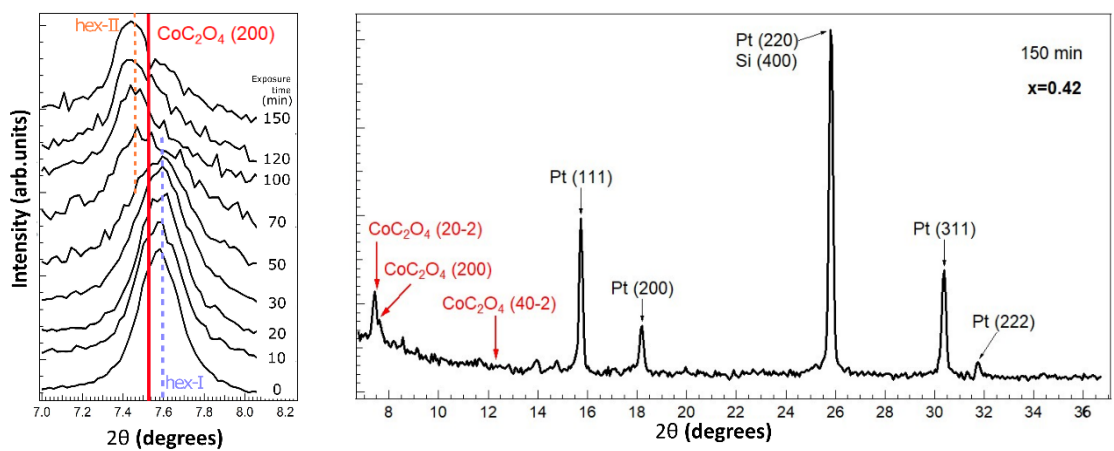

**Figure S4.** XRD spectra of a LCO thin film exposed to oxalic acid. The positions where the most intense cobalt oxalate peaks should appear are marked by red arrows. Peaks (200) and (20-2) coincide with LCO peaks, but the lack of intensity for the (40-2) reflection indicates that there is no cobalt oxalate present in significant amounts.

## 4) Changes in the sample morphology induced by chemical delithiation (by AFM)

We have also performed Atomic Force Microscopy images in several samples. In particular, Figure S5 shows the topography acquired in Amplitude Modulation Dynamic Mode of the two samples analyzed by XPS (Figure 9 in the main text). We can distinguish an increase in the roughness, in good agreement with the SEM analysis, and bundles of grains separated by deep grooves for the case of the delithiathed sample.

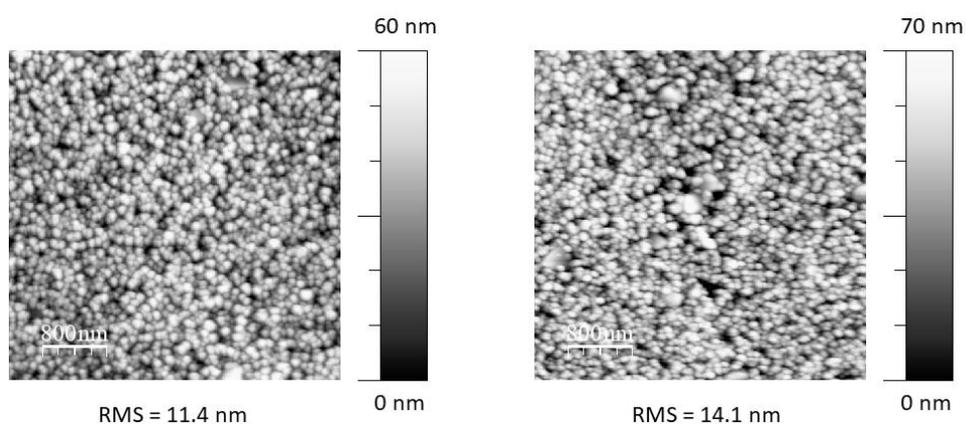

**Figure S5.** Change in the sample morphology with the chemical delithiation: (left) As-grown LCO film, and (right) 65 min-exposed at $10^{-2}$ M oxalic acid. Scanned area=2.5 x 2.5 µm² in Amplitude Modulation Dynamic Mode. The corresponding RMS roughness is specified at the bottom.